\documentclass{emulateapj}
\shorttitle{Search for Fallback Disks in Four Young SNRs}
\shortauthors{WANG ET AL.}

\newcommand{\chandra}{\textit{Chandra}}
\newcommand{\pup}{RX~J0822.0$-$4300}
\newcommand{\pks}{1E~1207.4$-$5209}
\newcommand{\rcw}{1E~161348$-$5055.1}
\newcommand{\cas}{CXOU~J232327.8+584842}

\defcitealias{pst04}{PST04}

\begin{document}
\bibliographystyle{apj_noskip}

%%\submitted{Accepted by ApJ}
\title{A Search for Fallback Disks in Four Young Supernova Remnants}

\author{Zhongxiang Wang, David L. Kaplan\altaffilmark{1,2}, and
Deepto~Chakrabarty\altaffilmark{2}} 
\affil{\small Kavli Institute for Astrophysics and Space
  Research, Massachusetts Institute of Technology, Cambridge, MA
  02139}  
\email{wangzx, dlk, deepto@space.mit.edu}

\altaffiltext{1}{Pappalardo Fellow}
\altaffiltext{2}{Also Department of Physics, Massachusetts Institute
  of Technology, Cambridge, MA 02139}

\begin{abstract}
We report on our search for the optical/infrared counterparts to the
central compact objects in four young supernova remnants: Pup A,
PKS 1209$-$52, RCW 103, and Cas A.  The X-ray point sources in
these supernova remnants are excellent targets for probing the
existence of supernova fallback disks, since 
irradiation of a disk by a central X-ray source should lead
to an infrared excess.  We used ground-based optical and near-infrared
imaging and {\em Spitzer Space Telescope} mid-infrared imaging
to search for optical/infrared counterparts at the
X-ray point source positions measured by
the {\em Chandra X-Ray Observatory}.  We did not detect any
counterparts, and hence find no evidence for fallback disks
around any of these sources.  In PKS 1209$-$52, we are able to
exclude a nearby optical/infrared candidate counterpart.  
In RCW 103, a blend of 3 faint stars at
the X-ray source position prevents us from deriving useful limits.  For
the other targets, the upper limits on the infrared/X-ray flux ratio
are as deep as (1.0--1.7)$\times 10^{-4}$.  Comparing these limits to
the ratio of $\approx 6\times10^{-5}$ measured for 4U~0142+61 (a young pulsar
recently found with an  X-ray irradiated dust disk), we conclude that
the non-detection of any disks around young neutron stars studied
here are consistent with their relatively low X-ray luminosities,
although we note that a similar dust disk around the neutron star in
Pup~A should be detectable by deeper infrared observations. 
\end{abstract}


\section{Introduction}

Neutron stars (NSs) are formed when massive stars end their lives
during core-collapse supernovae.  During such an explosion,
``fallback'' may occur when the reverse shock, caused by the impact of
%%%ejected core material and the stellar envelope, reaches the newly
the shock wave with the outer stellar envelope, reaches the newly
formed NS \citep{chevalier89, ww95}.  Depending upon
the rotation rate of the massive progenitor, some of the fallback may
have sufficient angular momentum to form a disk (\citealt{md81}; 
\citealt*{lwb91}).  Although widely predicted by contemporary
numerical studies of supernovae \citep{ww95}, there are
few observational constraints on the existence of fallback disks or
even of supernova fallback generally.  Fallback could have profound
implications for the endgame of massive star evolution, particularly
since it could lead a newborn NS to collapse into a black hole
\citep{chevalier89}.  Such an outcome might yet explain the lack of a
detectable pulsar in SN 1987A (\citealt{zam+98}; \citealt*{fcp99}). 
Supernova fallback is also a promising mechanism for
producing the debris disks necessary to form planets \citep{ph93, pod93},
which are known to exist around at
least one pulsar \citep{wf92, wol94}.

The compact stellar remnants of supernovae obviously probe supernova
physics.  While it was once thought that young radio pulsars like
those in the Crab and Vela supernova remnants (SNRs) were
prototypical of newborn NSs, it has recently been realized that the
young NS population is unexpectedly diverse.  For example, there are a
few young SNRs currently known to contain radio-quiet, non-plerionic
X-ray point sources near their geometric center (see \citealt*{pst04},
hereafter \citetalias{pst04}, for a review).  These X-ray point
sources, often called central compact objects (CCOs), have X-ray
spectra roughly consistent with the thermal-like surface (but smaller
than the expected neutron star surface) emission from young cooling
NSs.  However, they differ from ``typical'' young NSs (e.g., the Crab
pulsar) in SNRs in that they lack strong non-thermal emission
components and most of them do not have detectable pulsations.  Along
with other young NS classes including the anomalous X-ray pulsars
(AXPs), and soft gamma-ray repeaters \citep{wt04},
%, and dim isolated NSs
%(e.g., \citealt{haberl04}), 
these objects challenge our standard
framework for understanding young NSs.

Because CCOs and AXPs do not have the strong, broad-band, non-thermal
emission typically observed from young Crab-like pulsars and their
associated synchrotron nebulae, they offer an attractive opportunity
for testing the existence of fallback disks by detecting their
expected thermal emission.  \citet*{phn00} showed that the
combination of an untruncated outer edge of a fallback disk (in
contrast to a binary accretion disk) and irradiation of the disk by
the central pulsar should lead to a strong thermal excess in the
infrared and submillimeter bands.  In a previous paper, we reported
the detection of a mid-infrared counterpart to the AXP 4U 0142+61,
which we interpreted as dust emission from an X-ray heated debris disk
around the pulsar \citep*{wck06}.  Here, we
report our ground-based optical/near-IR and {\em Spitzer Space
Telescope} 4.5/8.0 $\mu$m observations of four CCOs. 
\begin{figure*}
%\plotone{f1.eps}
\includegraphics[scale=0.9]{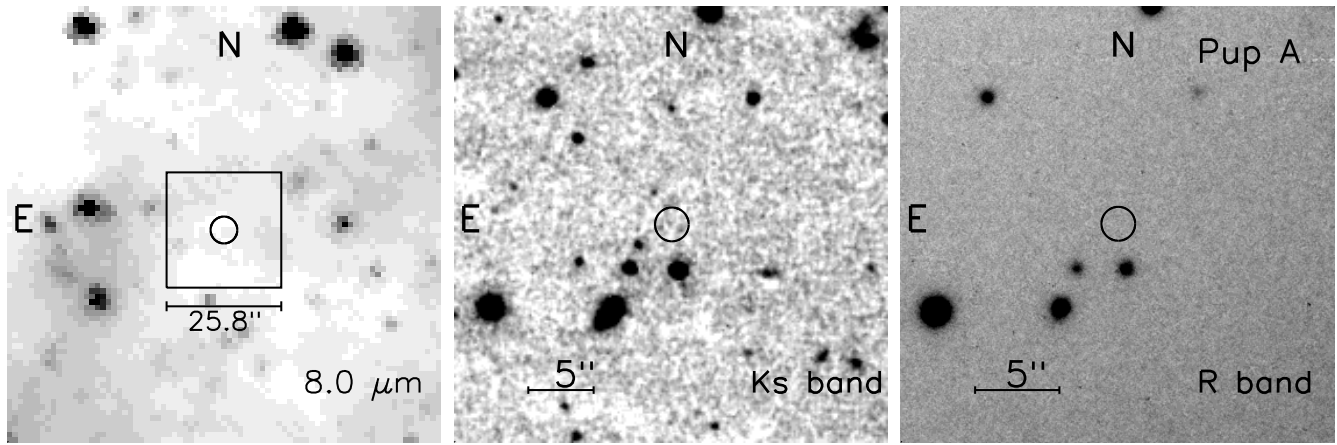}
\figcaption{Optical and IR images of \pup.  The {\em Chandra} error circle (solid circle) of the CCO
is indicated in the optical $R$ (right panel) and near-IR $K_s$
(middle panel) bands images.  The left panel shows the {\em
Spitzer}/IRAC 8.0 $\mu$m band image, with the box and circle %%%(not to scale)
indicating the size of the $R$ and $K_s$ images and the CCO position
respectively.  The circle is not to scale.  No object was found within the error circle.
\label{fig:pup}}
\end{figure*}
These targets were selected from a total of 8 known CCOs 
because they are relatively young and their distances are well determined
(\citetalias{pst04}).  
We give brief
summaries of our targets in \S~\ref{sec:targ}.  We then present the
optical/IR observations in \S~\ref{sec:obs}.  Finally, we present our
results in \S~\ref{sec:res} and our conclusions in \S~\ref{sec:disc}.
Part of our ground-based observations have
previously been reported \citep{wc02}.

%% We present the possible
%% identification of a near-IR counterpart to 1E~1614$-$5055 and the
%% identification of a near-IR object near the 1E~1207$-$5209 {\em
%% Chandra} error circle.  In the case of 1E~1614$-$5055, our counterpart
%% probably corresponds to three low-mass stars reported by PST04 (see
%% also Pavlov, Sanwal, Garmire 2002a).

\subsection{Target Summary}
\label{sec:targ}
Of the four CCOs that we observed, the basic properties of these
sources are similar: young, radio quiet, and X-ray bright, with
roughly similar X-ray spectra (again, see \citetalias{pst04}).
However, some have distinguishing characteristics, as we note below.
We also describe observations of their host SNRs used to determine
ages and distances, fundamental quantities that we use in the
interpretation of the observations.  A summary of the general
properties of the CCOs that we observed is given in
Table~\ref{tab:xps}.

\begin{description}
\item[\pup\ (Pup A)] The X-ray source \pup\ (\citealt{pkwc82};  \citealt*{pbw96}) is located in the
SNR Puppis~A (Pup A; G260.4$-$3.4).  The SNR
is the remnant of a Type II explosion that occurred $<5000$~yrs ago
(\citealt{wtki88};  \citealt*{adp91}).  The distance to the SNR, determined from
associated \ion{H}{1} absorption, is 2.2$\pm0.3$~kpc \citep{rgj+03}.
The CCO is unremarkable, except that \citet{hb05} recently reported a
possible detection of 0.22-s X-ray pulsations.

\item[{\pks} (PKS 1209--52)] The source \pks\ \citep{hb84} is in the SNR PKS~1209$-$52
(G296.5+10.0).  This SNR has a distance of
2.1$^{+1.8}_{-0.8}$~kpc based on  associated \ion{H}{1} gas
\citep{gbg+00} and an age of $\approx 7$~kyr \citep{rmk+88}. The CCO
is the only confirmed pulsator among the sources considered here
($P_{\rm spin}$=424 ms; \citealt{zpst00,pzst02}) although its spin
does not change monotonically \citep*{zps04}, and its X-ray spectrum
is quite unique with a number of related X-ray absorption features
\citep{spzt02,bcdlm03}.

\item[{\rcw} (RCW~103)] The CCO \rcw\ \citep{tg80} is in the SNR RCW~103
(G332.4$-$0.4).  The distance to the SNR of
3.1~kpc has been estimated from an \ion{H}{1} association
\citep{rgj+04}, while the age of $\approx 2$~kyr was determined from
measurement of the expansion of optical filaments \citep*{cdb97}.
Unlike the other objects, the CCO has shown considerable X-ray
variability on a timescale of years \citep*{gpv99,ba02}, and at least
three faint near-infrared sources have been found within the 0\farcs7
{\em Chandra} X-ray position uncertainty (\citetalias{pst04}).
Together, these facts suggest that \rcw\ is a low-mass X-ray
binary with a 6.4 hr orbital period (\citealt{ba02};
\citetalias{pst04}).

\item[{\cas} (Cas~A)] \cas\ \citep{t99} is in the SNR
Cassiopeia A (Cas A; G111.7-2.1), the
youngest known Galactic core-collapse SNR ($\approx$300 yr;
\citealt*{tfvdb01}), but the CCO was only revealed recently (compared
to the others, which had been identified by the {\em Einstein
Observatory}) by the {\em Chandra} X-ray Observatory
\citep{t99,pza+00,cph+01}.  Comparing proper motions and radial
velocities of ejecta, \citet{rhfw95} estimate a distance of
$3.4_{-0.1}^{+0.3}$~kpc.  The X-ray properties are, like those of
\pup, unremarkable, and this CCO serves as a benchmark for
the class. Deep optical/near-IR observations of the source field were
conducted \citep*{kkm01,fps06}, but no counterpart was found.
\end{description}

%% \begin{figure}
%% \centerline{\includegraphics[width=\textwidth]{pupam.eps}}
%% \centerline{\includegraphics[width=\textwidth]{pksm.eps}}
%% \centerline{\includegraphics[width=\textwidth]{rcwm.eps}}
%% \caption{Optical and IR images of RX J0822--4300 (\textit{top}), 1E
%% 1207.4$-$5209 (\textit{middle}), and 1E 1614$-$5055 (\textit{bottom}).
%% The left panels show the {\em Spitzer}/IRAC 8.0 $\mu$m  images of
%% the fields with the boxes indicating the sizes of the $K_s$ and $i'$
%% images, shown in the middle and right panel respectively.  The circles
%% indicate the CCO position uncertainties.}
%% \end{figure}

\begin{figure*}
%fig2
%%\plotone{f2.eps}
\includegraphics[scale=0.9]{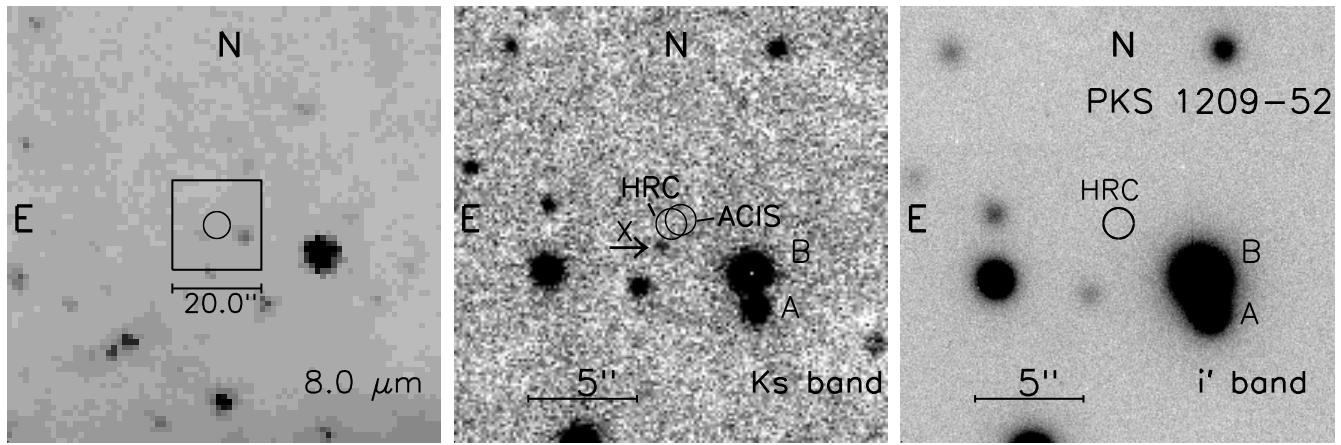}
\figcaption{Optical/IR images of the 1E
1207.4$-$5209 region.  In the left panel, the {\em Spitzer}/IRAC 8.0
$\mu$m band image is shown, with the box and circle %%%%(not to scale)
indicating the size of the middle and right panel images and the CCO
position respectively.  The circle is not to scale. 
In the middle panel, a $K_s$ band image is
shown. The two solid circles indicate the {\em Chandra} HRC and ACIS
positions, which are about 1\farcs0 away from the object labeled {\em
X}.  The right panel shows the optical $i'$ band image of the target
field.  The near-IR object was not detected by our optical
observations, but was found with $V\simeq 26.8$ by Pavlov et al.\
(private communication).  Previous observations have identified the
two blended stars ($A$ and $B$) as main sequence stars
(\citealt*{mlt88}; \citealt*{bcm92}; see also \citealt{dlmc+04}).
\label{fig:pks}}
%\end{center}
\end{figure*}

\section{Observations}
\label{sec:obs}

\subsection{Ground-based Observations}
%\subsection{Magellan Optical Observations}
We obtained ground-based observations of the three southern targets
among our sample of four CCOs.  A summary of these observations is
given in Table~\ref{tab:obs}. Below we briefly list the observations,
and discuss the reduction and calibration procedures. 

We obtained optical images of the fields around the CCOs \pup, \pks,
and \rcw\ using the Magellan Instant Camera (MagIC) at an
f/11 Nasmyth focus of both the 6.5~m Baade (Magellan~I) and Clay
(Magellan~II) telescopes \citep{sj03} at Las Campanas Observatory in
Chile.  MagIC is a 2048$\times$2048 SITe CCD with a
0\farcs069~pixel$^{-1}$ plate scale and a 142\arcsec\ field of view.
We observed the standard stars SA~107--614 and Ru~152 \citep{lan92,
smi+02} to calibrate the Baade data, while for the Clay data we also
observed a number of
Stetson\footnote{\url{http://cadcwww.dao.nrc.ca/cadcbin/wdbi.cgi/astrocat/stetson/query}}
standard fields with both MagIC and the Las~Campanas 40-inch
telescope.  We reduced the MagIC data using the {\tt IRAF} data
analysis package, first subtracting bias images and then applying
flat-fields.

%\subsection{CTIO/Magellan Near-IR Observations}

We obtained $JHK_s$ near-IR images of the \pup\ and
\pks\ fields using the Ohio State Infrared
Imager/Spectrometer (OSIRIS; \citealt{dab+93}) at the f/14 tip-tilt
focus of the 4~m Blanco Telescope at the Cerro Tololo Inter-American
Observatory (CTIO) in Chile.  The detector in OSIRIS was a Rockwell
HAWAII HgCdTe 1024$\times$1024 array.  We used the OSIRIS f/7 camera,
which had a 0\farcs161~pixel$^{-1}$ plate scale and a 93\arcsec\ field
of view.  During each exposure, we dithered the telescope in a
3$\times$3 grid with offsets of about 10\arcsec\ to allow for
correction of the rapidly variable infrared sky background and to
minimize the effect of bad pixels.  The near-IR standard stars SJ~9136
(HST S165-E) and SJ~9157 (HST S273-E) were observed for flux
calibration \citep{pmk+98}.

We then obtained $JHK_s$ images of our target fields with the
``ClassicCam'' near-infrared imager \citep{pwc+92} at an f/11
Nasmyth focus on the Baade Telescope.  The detector was a Rockwell
NICMOS3 HgCdTe 256$\times$256 array.  We operated the camera in its
low-resolution mode, with a 0\farcs112~pixel$^{-1}$ plate scale and a
29\arcsec\ field of view.  The same dithering and data reduction was
employed as with the CTIO observations.  The near-IR standard star
SJ~9172 (HST S279-F) was observed \citep{pmk+98}.

Finally, we obtained deep near-IR images of the \pks\ field
under excellent conditions using the Baade telescope and Persson's
Auxiliary Nasmyth Infrared Camera (PANIC; \citealt{mpm+04}), which
contains a Rockwell 1024$\times$1024 detector with a 0.125\arcsec\
plate scale.  The telescope position was dithered five times with a
15\arcsec\ dither step, over the course of each exposure.  The
standard star SJ~9172 was also observed.

As with the optical data, the {\tt IRAF} data analysis package was
used for the data reduction, including bias subtraction and flat
fielding. From each set of the dithered near-IR images, a sky image
was derived by filtering out stars. The sky image was then subtracted
from the set of images, and a final target field image was obtained by
average combining the sky-subtracted images.

We astrometrically calibrated all of our ground-based optical and
near-IR images by matching unsaturated field stars to USNO-A2.0
astrometric catalog stars \citep{m98}.  We used 24, 46, and 50 stars in
calibrating \pup, \pks, and \rcw\
respectively. We used the {\tt IRAF} task {\tt ccmap} to derive the
astrometric solutions. The uncertainties in matching the stars are
negligible.  The nominal uncertainties of the calibrated images are
dominated by the USNO-A2.0 catalog systematic uncertainty 
($\simeq$0\farcs25; \citealt{m98}). 

\subsection{{\em Spitzer}/IRAC Mid-IR Observations}

We observed the four CCOs with the \textit{Spitzer Space Telescope}
using the Infrared Array Camera (IRAC; \citealt{fha+04}), an imaging
camera operating at wavelengths of 3.6, 4.5, 5.8, and 8.0 $\mu$m. One
pair of the detectors of the IRAC, either 3.6/5.8 $\mu$m or 4.5/8.0
$\mu$m, observes one field of view simultaneously: we used the 4.5/8.0
$\mu$m detectors. The detectors at the short and long wavelengths are
InSb and Si:As devices respectively, with 256$\times$256 pixels and a
plate scale of 1\farcs2, giving a field of view of
5\farcm2$\times$5\farcm2.  The frame time we used was 100 s, with 96.8
s effective exposure time per frame for the 4.5 $\mu$m data and 93.6 s
effective exposure time for the 8.0 $\mu$m data.

The data were processed through the data reduction pipelines (version
S11.0; IRAC Data Handbook 2004) at the {\em Spitzer} Science
Center. In the basic calibrated data (BCD) pipeline, the individual
flux-calibrated frames were produced from the raw images. The BCD
\begin{figure*}
%fig3
%%\plotone{f3.eps}
\includegraphics[scale=0.9]{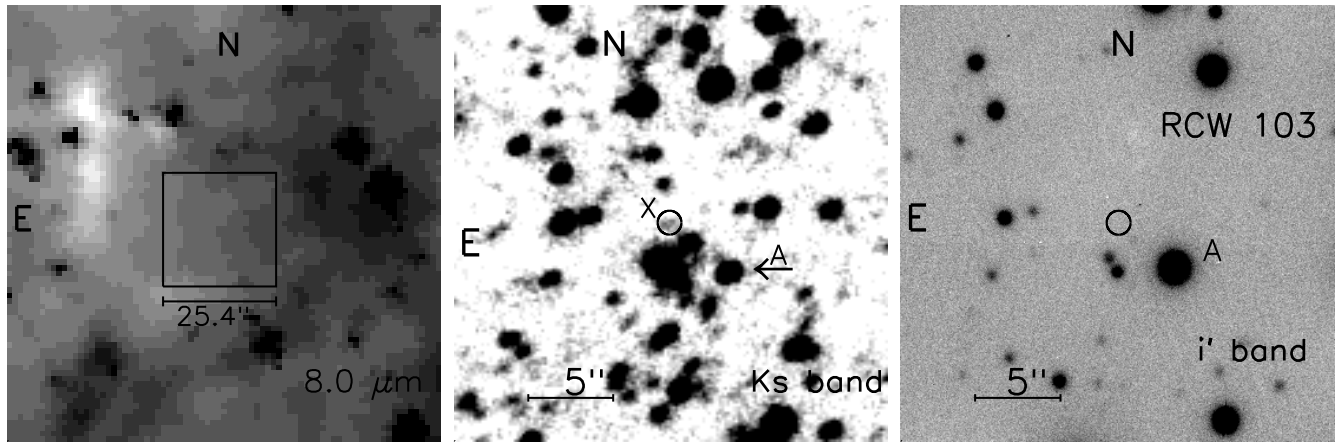}
\figcaption{Optical/IR images of the region
around \rcw. The left panel shows the {\em Spitzer}/IRAC 8.0
$\mu$m band image of the field with the box indicating the size of the
$K_s$ and $i'$ images, shown in the middle and right panel
respectively. The CCO position is at the center of the box, and no
object was found.  One object, probably consisting three faint objects
as reported by \citetalias{pst04}, was detected by the near-IR
observations within the 0\farcs7 {\em Chandra} error circle (middle
panel). The object was not detected by our optical observations, as
indicated by the right panel image.  Object A was identified as a
subdwarf M star by \citet{tdgm83}.\label{fig:rcw}}
%\end{center}
\end{figure*}
frames were then average combined into the post-BCD (PBCD) mosaics. In
the PBCD pipeline, each frame was astrometrically calibrated with
respective to the detected Two Micron All-Sky Survey (2MASS;
\citealt{2mass}) stars, resulting in a position
accuracy\footnote{See
\url{http://www.ipac.caltech.edu/2mass/releases/allsky/doc/explsup.html}.} of
$<$0\farcs3.

Our IRAC observations did not detect any mid-IR counterparts to the
CCOs. The sensitivity of IRAC observations is dominated by a
combination of background sky emission and confusion, with the latter
dominating when a field is crowded.  We measured the IRAC sky
brightnesses at our target positions (\S~\ref{sec:ast}), and we give
those values in Table~\ref{tab:sky}.  As can be seen, Pup A and PKS
1209-52 have low background emission while Cas A has medium/high
background emission\footnote{See
\url{http://www.spitzer.caltech.edu/obs/bg.html} for a discussion of
{\em Spitzer} background levels}, caused somewhat by diffuse emission
from the SNR; the high background of RCW 103 is presumably caused by
its crowded field.  To determine upper limits to the presences of
point sources in the IRAC images, we converted the PBCD images of our
targets in units of MJy/sr to data numbers (DN), and derived the
3$\sigma$ upper limits (given in Table~\ref{tab:lims}) from the
standard deviation of the background sky at the source positions.  

\subsection{X-ray Positions}
\label{sec:ast}
The X-ray positions of the CCOs reported in the literature were not of
the uniformly high quality necessary for comparison with ground-based
data.  Where possible we took positions from published {\em Chandra
X-Ray Observatory} results, but for three of the four sources we
derived new positions from one or more observations retrieved from the
{\em Chandra} public data archive.  Using the CIAO thread {\tt
fix\_offset}, we checked the data for any known aspect offsets and
applied the corrections if necessary. The CIAO tool {\tt wavdetect}
was used for obtaining the source positions.  We give all of the
positions in Table~\ref{tab:xps}, and discuss the observations of each
source in more detail below.  The total nominal position uncertainty
for locating the CCOs on our optical/IR images is dominated by
the \chandra\ absolute astrometric uncertainty ($\approx$0\farcs6
with a 90\% confidence). To estimate the effects of relative systematic
astrometric errors between the optical/IR and \chandra\ images, 
we adopted the approach outlined in \S~\ref{subsec:pks}.

%summing in quadrature the {\em Chandra} absolute astrometric
%uncertainty\footnote{See
%\url{http://cxc.harvard.edu/cal/ASPECT/celmon/}.} ($\simeq$0\farcs6
%with a 90\% confidence, or $\simeq0\farcs35$ at 1$\sigma$) and the
%USNO-A2.0 catalog systematic uncertainty ($\approx 0\farcs25$), giving
%a 1$\sigma$ radius of $\approx 0\farcs43$.

\section{Results}
\label{sec:res}

\subsection{\pup\ (Pup A)} 
For \pup\ there are three {\em Chandra} observations made with the HRC-I
(Obs\_ID=749, exposure-time=18 ks),
ACIS-S (Obs\_ID=750, exposure-time=12 ks), and HRC-S
(Obs\_ID=1851, exposure-time=20 ks).
The source positions we derived from the data generally differ by
0\farcs2, consistent with the {\em Chandra} pointing uncertainty
($\simeq$0\farcs6 with a 90\% confidence). We adopted the position
from ACIS-S data: R.A.~= 08$^{\rm h}$21$^{\rm m}$57\fs42 and Dec.~=
$-$43\arcdeg 00\arcmin 16\farcs62 (equinox J2000.0); ACIS-S positions
have been tested more and should be more reliable \citep{fps06}.  The
total nominal uncertainty of the position on the optical/near-IR images is
0\farcs7 (90\% confidence), dominated by the {\em Chandra} and
USNO-A2.0 catalog systematic uncertainties.  In Figure~\ref{fig:pup}
we plot optical, near-, and mid-IR images of \pup\ and indicate the
X-ray position as a solid circle in the middle and right panels of the
figure.  No objects were found near the position of \pup, so we
determined the 3$\sigma$ limiting magnitudes from our observations and
give them in Table~\ref{tab:lims}.

\subsection{\pks\ (PKS 1209--52)}
\label{subsec:pks}
The source \pks\ was observed with {\em Chandra}/ACIS
several times between 2000 and 2003, but only the 2003 June~6
observation (Obs\_ID=3913,
exposure-time=20 ks) with ACIS-S was not under continuous clocking
(CC) mode (which only allows derivation of one-dimensional source
positions).  From this observation we derive a position of R.A.~=
12$^{\rm h}$10$^{\rm m}00\fs92$ and Dec.~=
$-$52\arcdeg26\arcmin28\farcs35\ (equinox J2000.0).  In addition,
there is an HRC-I observation made on 2003 December~28
(Obs\_ID=4593, exposure time=50 ks)
which gives the position R.A.~= 12$^{\rm h}$10$^{\rm m}00\fs88$ and
Dec.~= $-$52\arcdeg26\arcmin28\farcs66\ (equinox J2000.0), differing
by about 0\farcs5 from the ACIS position, with the uncertainty on each
of about 0\farcs7 (90\% confidence). 
In Figure~\ref{fig:pks}, we indicate both ACIS and
HRC positions; while the ACIS position should be more reliable (see
above), since there is a source near the error circle of this CCO (see
below) we show both positions to demonstrate that 
%%the position difference does not depend on choice of instrument.
neither of them is consistent with the source.

As shown by the $K_s$ image in Figure~\ref{fig:pks}, we detected one
faint object (labeled {\em X}) in our $J$ and $K_s$ near-IR images. We
used the {\tt IRAF} aperture photometry package {\tt APPHOT} to
measure the brightness of this object, finding $J=22.2 \pm 0.3$ and
$K_s = 21.1\pm0.2$.  The detection of this object was also reported by
\citetalias{pst04} ($V\simeq 26.8$, $K_s\simeq 20.7$), and it was
suggested as a low mass dwarf (G. Pavlov, private communication).
This object is the closest to the X-ray error circle, but is 1\farcs06
away from the HRC position and 1\farcs36 away from the ACIS position:
both offsets are close to
%%larger than expected from 
the positional uncertainty limit, 
with deviations of $\approx3\sigma$ from the ACIS
position and $\approx 2.5\sigma$ from the HRC position.
%%However, to assess the reality of the association even further, we
Our further examination of the ACIS image found three X-ray sources 
coincident with stars from the USNO-A2.0 catalog.  These stars, having
7--200 X-ray counts, were all within $3\arcmin$
of the ACIS aimpoint such that the astrometry should be of the
highest quality\footnote{Again, see
  \url{http://cxc.harvard.edu/cal/ASPECT/celmon/}.}. 
%have varying numbers of X-ray counts (7--200) but are undoubtedly
%real X-ray sources.  
The X-ray positions that we measured for the three
stars differ by an average of $0\farcs11\pm 0\farcs08$ in R.A. and
$0\farcs34\pm 0\farcs08$ in Dec. from the USNO positions, 
indicating that we would expect at most a $0\farcs7$ offset between X-ray and
USNO positions (at 3$\sigma$ confidence), which means that the
$>1\arcsec$ offset between {\em X} and \pks\ is highly unlikely.
%% : both offsets are larger than the astrometric uncertainty that is
%% dominated by the well-tested {\em Chandra} pointing accuracy.
%% Although the object is within the 3$\sigma$ error circle (radius
%% 1\farcs35) of the HRC position, in our opinion, the position
%% discrepancy is too large for the association to be real.  
As an
independent check, our photometry indicates that the object is likely
to be a background main-sequence star (M5 dwarf at a distance
of 8 kpc). We therefore dismiss this
source as an unrelated star.  Aside from this object, we detect no
likely counterpart to the CCO, and so we give 3$\sigma$ limiting
magnitudes in Table~\ref{tab:lims}.
%\begin{figure}
\begin{center}
%%\plotone{f4.eps}
\includegraphics[scale=0.75]{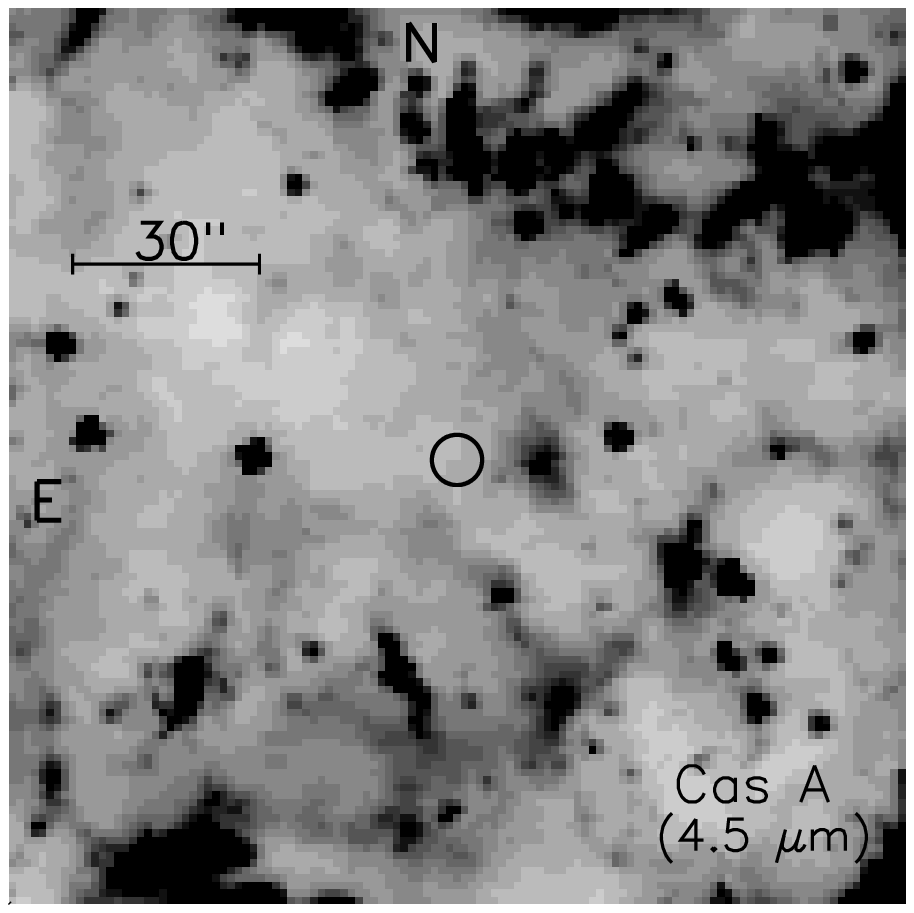}
\figcaption{{\em Spitzer}/IRAC 4.5 $\mu$m band
image of Cas A. The CCO position is indicated by a 4\farcs0 radius
circle (this is far larger than the position uncertainty, and is
intended for ease of viewing). No counterpart was
found.\label{fig:cas}}
\end{center}
%%\end{figure}

\subsection{\rcw\ (RCW~103)}
The CCO \rcw\ has been monitored over the last few years with
{\em Chandra}/ACIS, and most of the observations were short with
3.5--5 ks exposures.  We analyzed the data taken on 1999 September 26
(Obs\_ID=123, exposure-time=14 ks),
%data analysis package.\footnote{See http://asc.harvard.edu/ciao/}
and obtained the following position of \rcw: R.A.~= 16$^{\rm
h}$17$^{\rm m}36\fs25$ and Dec.~= $-$51\arcdeg02\arcmin24\farcs5\
(equinox J2000.0), which is consistent with that reported by
\citet{gpgz00}.  In Figures~\ref{fig:rcw}, we show the source position
on our $i'$ and $K_s$ images of the field of \rcw.  For
\rcw, the total nominal uncertainty (90\% confidence) on the 
optical/IR images, taking into
account the Chandra pointing and USNO-A2.0 calibration uncertainties,
is 0\farcs7.

One object (labeled {\em X} in the middle panel of
Figure~\ref{fig:rcw}) was detected at the {\em Chandra} position in
our near-IR $HK_s$ images.  We used the {\tt IRAF} photometry package
{\tt DAOPHOT} to measure the magnitudes of the stars in this crowded
field, and found $H = 19.4\pm 0.2$ and $K_s = 18.1\pm 0.1$, along with
a 3$\sigma$ limiting magnitude of $J\geq 21.9$.  We believe that this
object is the same as one reported by \citet{psgz02}, who measured
$J\simeq 22.3$. \citetalias{pst04} subsequently used high-resolution
ground-based and \textit{Hubble Space Telescope} imaging to identify
three faint late-type stars within the 0\farcs7 {\em Chandra} error
circle.  Our object {\em X} probably corresponds to the blend of
these.  PST04 have suggested that \rcw\ may be an X-ray
binary with a late-type companion.  In bands other than $H$ and $K_s$, we
do not detect any counterpart, and hence give 3$\sigma$ limiting
magnitudes in Table~\ref{tab:lims}.  The limits are not as deep as
those of the previous two sources because of higher background in this
field.

\subsection{\cas\ (Cas A)} We adopted the position given by
\citet{fps06} for \cas, which is R.A.~= 23$^{\rm
h}$23$^{\rm m}$27\fs94, Dec.~= +58\arcdeg 48\arcmin 42\farcs51
(equinox J2000.0), with a 1$\sigma$ uncertainty of 0\farcs4. This
position is indicated in Figure~\ref{fig:cas}, where we display the
4.5~$\mu$m IRAC image of the field.  No mid-IR counterpart was
detected in our data, so we give 3$\sigma$ limiting magnitudes in
Table~\ref{tab:lims}.  The limits are considerably less constraining
than those of \pup\ because of the high background from
the SNR.

\section{Discussion and Conclusions}
\label{sec:disc}
Based on the current observed properties, it is likely that the CCOs
represent elements of a new class of young NSs, or even of several new
classes (given their heterogeneous properties). Their overall
characteristics --- radio-quietness, distinct thermal X-ray emission,
relatively long spin periods (when detected) for their ages, and lack
of pulsar wind nebulae --- indicate that they distinctively differ
from other classes of young NSs, especially the young rotation-powered
pulsars.  The possible origin of the CCOs has been well discussed by
comparing them to different types of NS objects and models, which
include the rotation-powered pulsars, AXPs, magnetars, accreting
binaries, accreting NSs from a residual disk (e.g., \citealt{cph+01};
\citetalias{pst04}; \citealt{ghs05}).  Among them, one intriguing
model is that the CCOs are NSs accreting from a residual disk,
probably formed from fallback of supernova material.  This model was
primarily proposed as an alternative to the magnetar model for AXPs in
order to explain their ``anomalous'' natures: that their X-ray
luminosities are much higher than their rotational energy loss rates
(e.g., \citealt*{chn00}; \citealt{mlrh01}).  However, the X-ray fluxes
of most of the CCOs, especially our targets (e.g.,
\citealt{zps04,fps06}), have been steady based on the observations
over the past few years, contrary to the fact that accreting X-ray
sources are highly variable (this is in contrast to \rcw, which
\citetalias{pst04} therefore exclude from the CCO category).  In
addition, the optical observations of CCOs have failed to detect any
counterparts that could be an accretion disk (e.g.,
\citealt{fps06,ghs05}). Particularly for \pks, which is the least
reddened among the CCOs, the optical observations (including ours)
%%(see Figure~\ref{fig:all}), 
were deep enough to exclude the existence of an
accretion disk around it (except an edge-on disk, which has very low
probability; \citealt{zps04}).  Therefore, even though the CCOs are
probably not a uniform class of objects, it is very unlikely that
their X-ray emission is powered by accretion.
%Our optical observations of 1E 1207.4$-$5209, with the upper limits
%about 24~mag, have provided additional evidence for excluding the
%existence of an accretion disk (except an edge-on disk, which has very
%low probability; Zavlin et al. 2004) in this source.  
%The neutron
%star's surface magnetic field is $B_s\simeq 3\times 10^{12}$ G,
%assuming its spin period $P=0.424$ s and period time derivative
%$\dot{P}=2\times 10^{-14}$ s s$^{-1}$ \textbf{DLK: This $\dot P$ is not
%true}.  The optical flux from an accretion disk, which is sensitive to
%the inner disk radius and thus determined by the surface magnetic
%field for a given mass accretion rate, would have $V\lesssim 23$
%(Zavlin et al. 2004).  
%To illustrate this, we show the observed
%ultraviolet/optical fluxes from an accretion disk in the ultra-compact
%binary 4U 1626$-$67 (e.g., Chakrabarty 1998) in Figure 5. The fluxes,
%scaled by the X-ray flux ratio between 1E 1207.4$-$5209 and 4U 1626$-$67
%(here we used its 0.4--10 keV X-ray flux $F_{\rm X}=3 \times 10^{-10}$
%ergs s$^{-1}$ cm$^{-2}$, derived from the {\em ASCA} data), would be
%detectable by our optical observations.  
%For RX J0822.0$-$4300, it is
%unclear about its surface magnetic field.  Its recently reported $P$
%($0.22$ s) and $\dot{P}$ ($2.1\times 10^{-10}$ s s$^{-1}$; Hui \&
%Becker 2005) suggest an extremely strong surface magnetic field
%$B_{s}$= 2$\times 10^{14}$ G.
\begin{figure*}[t]
\begin{center}
%%\vspace*{-5mm}
%%\plotone{f5.eps}
%fig5
\includegraphics[scale=0.80]{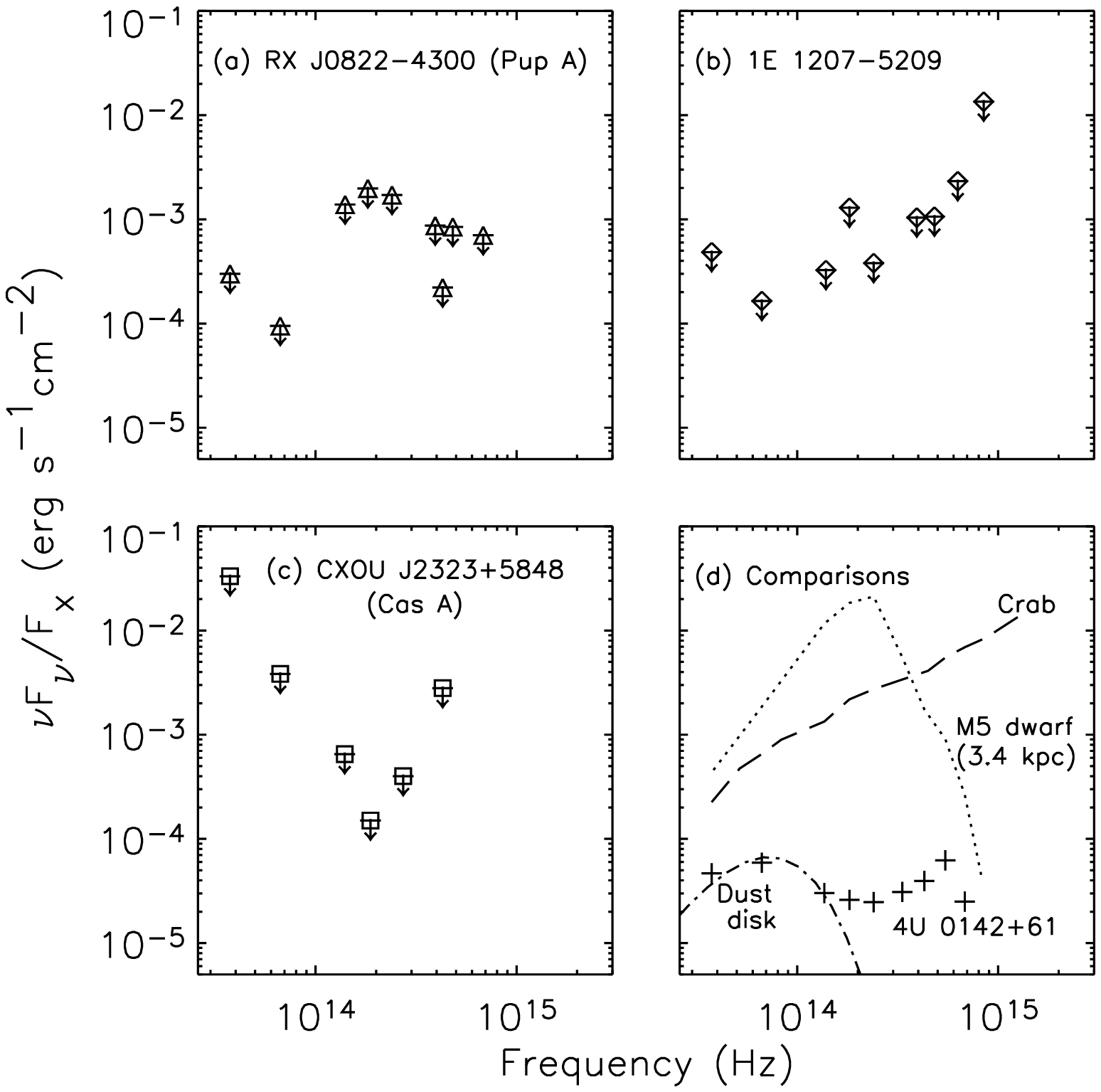}
\figcaption{Upper limits on the optical/IR to X-ray flux ratio
for the CCOs: ($a$) \pup\ ; ($b$) \pks\ ; ($c$) \cas\ 
(the optical/near-IR $RJHK$ upper limits are from \citealt{fps06}). 
Fluxes are unabsorbed, and 
$A_V=$1.6, 0.5, and 7 are respectively used 
for dereddening the optical/IR data (according to
the reddening laws of \citealt*{sfd98} for the $BVRI$ data,
\citealt{fan99} for the $u^{\prime}g^{\prime}r^{\prime}i^{\prime}$
data, \citealt{rl85} for the $JHK$ data, and \citealt{imb+05} for the
IRAC data).
%%The limits of each CCO are normalized by its unabsorbed X-ray flux 
%%(\citetalias{pst04}).  
($d$) Optical/IR to X-ray flux ratios (for comparision) of 
the AXP 4U 0142+61 (plus sign; \citealt{wck06}) and
the Crab pulsar (long-dashed curve; \citealt{efr+97}; Temim et al.\ 2006,
in preparation).  The IR component of 4U 0142+61
is likely to be due to an X-ray irradiated dust disk (dash-dotted
curve). The optical/IR observations should also
have detected a main-sequence star down to M5 (dotted curve; 
from \citealt{allen} and \citealt{kurucz93}, and normalized to a 
typical 0.4--8.0 keV CCO X-ray flux 2$\times 10^{-12}$ erg cm$^{-2}$ s$^{-1}$)
even at a distance of 3.4 kpc. 
\label{fig:all}
}
\end{center}
\end{figure*}

However, it is still possible that isolated, young NSs have cool,
passive disks. Such a disk would be farther away from a central NS and
cooler than an accretion disk.  
%Since there is no binary companion to
%truncate the outer edge of the disk, the outer portion of the disk
%would extend to regions where the material is cooler than 1000~K and
%dust grains would form and produce detectable IR emission.  
If they exist, the disks would not contribute to generating X-ray
emission from young NSs such as the CCOs and AXPs, but their existence
might help explain various phenomena of radio pulsars
\citep{md81,bp04,cs06}, planets formation around NSs \citep{lwb91}, or
the diversity of young NSs \citep{alpar01}.  The recent discovery of
the mid-IR emission from the AXP 4U~0142+61 probably indicates the
existence of such a passive disk around that young pulsar
\citep{wck06}.  The disk, irradiated by the X-rays from the central
X-ray pulsar, emits mainly in the IR (Figure~\ref{fig:all}).  Using
{\em Spitzer}/IRAC observations similar to those discussed here, its
unabsorbed IR--to--X-ray flux ratio was found to have a maximum value
of $(\nu_{4.5\mu}F_{4.5\mu})/F_{\rm X} \simeq$6.3$\times 10^{-5}$ at
4.5 $\mu$m.  This ratio, as it depends largely on the geometry of the
disk and not on the peculiarities of the central object, might be
typical for any dust disks around young NSs. In Figure~\ref{fig:all},
we compare the upper limits on the IR/X-ray flux ratios of the CCOs to
the same ratios of 4U~0142+61.  
%As can be seen, the non-detection of
%the sources in the IR could be due to their relatively low X-ray
%luminosities ($\simeq$3$\times$10$^{33}$ ergs s$^{-1}$; see Table~1).
The upper limits on the IR/X-ray flux ratios of the CCOs are not 
sufficiently deep to detect a disk similar to that found in 4U 0142+61. 
As the X-ray luminosities of the CCOs are typically lower 
($\simeq$$2\times 10^{33}$ erg s$^{-1}$; see Table 1), deeper IR exposures are needed in order 
to reach a comparable IR/X-ray flux ratio.
Accordingly, in order to further explore any existence of dust disks
around the CCOs, \pup\ would be a plausible target. For
example, a 4.5 $\mu$m flux of 0.6 $\mu$Jy would be expected from a
dust disk, which is detectable by deep {\em Spitzer}/IRAC observations
(several hours exposure time).  In comparison, for the very young
CCO \cas, although mid-IR observations are less affected by
reddening ($A_V\simeq 7$; \citealt{fps06}), diffuse emission of the
SNR highly contaminates the source field, making it extremely
difficult to have a sensitive search.

It is interesting to compare the flux ratio limits to those of the
Crab-like, rotation-powered pulsars, even though it is clear that
their optical/IR emission is nonthermal, originating from the
magnetospheres of the pulsars.  For several well-studied
rotation-powered pulsars, optical/IR emission can be connected to
nonthermal hard X-rays with a same power-law (e.g., PSR B0656+14;
\citealt*{pzs02}).  The connection may be further indicated by a
similar optical to nonthermal--X-ray flux ratio, $F_{\rm opt}/F^{\rm
nonth}_{\rm X}\simeq 0.001$--0.01, where $F_{\rm opt}$ is the optical
flux between 4000--9000 \AA\ and $F^{\rm nonth}_{\rm X}$ is the
non-thermal X-ray flux between 1--10 keV \citep{zp04}.  In
Figure~\ref{fig:all}, we also show the unabsorbed SED of the Crab
pulsar, normalized by its X-ray flux (1.5$\times$10$^{36}$ ergs
s$^{-1}$; \citealt{zp04}).
%%%%%$(\nu_{4.5\mu}F_{4.5\mu})/F_{\rm X}$
%$F_{4.5 \mu}\simeq 3.1$ mJy for the Crab pulsar
The SED is above most of the upper limits on the CCOs, indicating that
our observations would have detected Crab-like young pulsars.  This
comparison has additionally shown that the CCOs are a different class
of young NSs from the typical Crab-like pulsars.

Our optical/IR observations have additionally provided strong
constraints on the binary model proposed for the CCOs.  As the upper
limits of our observations of \pup\ and \pks\ are
24--26 mag in the optical and 20--23 mag in the near-IR (or $\nu
F_{\nu}\leq 5\times 10^{-15}\mbox{ erg s}^{-1}\mbox{ cm}^{-2}$; see
Figure~\ref{fig:all} and note that the X-ray fluxes are around
10$^{-12}\mbox{ erg s}^{-1}\mbox{ cm}^{-2}$), they would have detected
main-sequence stars down to spectral type M5 regardless of extinction
or white dwarfs down to effective temperature $T_{\rm eff}\simeq$
10$^4$ K \citep*{bwb95}, assuming relatively low extinction (e.g.,
$A_V\leq$1--2; \citealt{ps95}; see Table~1), out to distances of 2--3~kpc.
%% However, as discussed in \S~\ref{subsec:pks}, we can not rule out 
%% a very low-mass star or
%% a brown dwarf companion ($M\leq 0.1 M_{\sun}$).
%With such a low mass star, 1E 1207.4$-$5209 might be a wide binary
%(orbital period $P_{\rm orb}\sim 0.2$--6 yr), which may explain 
%the variations detected in its spin
%period (Zavlin et al. 2004).

In summary, we have obtained  deep multi-wavelength observations of
four CCOs in young SNRs. The observations have allowed us to find a
near-IR counterpart to the CCO in RCW 103, which probably corresponds
to three faint late type stars resolved by  high-resolution
observations.  In addition, the observations have detected a low mass
star close to, but inconsistent with, the current {\em Chandra}
positions of \pks.
%The star would be a 0.055--0.070
%$M_{\sun}$ brown dwarf at 1E 1207.4$-$5209's distance.
%, consistent with
%the wide binary scenario proposed to explain the variations detected
%in its spin period \citep{zps04}.  
The non-detection of any IR emission from debris disks around three of
the CCOs could be due to their relatively low X-ray luminosities,
compared to 4U~0142+61.  While high background prohibits additional
observations of two of the sources, deeper IR observations of \pup\
could probe the existence of a fallback disk around this CCO.

\acknowledgements IRAF is distributed by the National Optical
Astronomy Observatories, which are operated by the Association of
Universities for Research in Astronomy, Inc., under cooperative
agreement with the National Science Foundation.  The Digitized Sky
Surveys were produced at the Space Telescope Science Institute under
U.S. Government grant NAG W-2166.  This research has made use of the
data products from the Two Micron All Sky Survey, which is a joint
project of the University of Massachusetts and the Infrared Processing
and Analysis Center/Caltech, funded by NASA and NSF.

{\it Facilities:} Spitzer (IRAC), Magellan:Baade
  (ClassicCam,MagIC,PANIC), Magellan:Clay (MagIC), Blanco (OSIRIS)

\bibliographystyle{apj}
%\bibliography{msdlk}

%%\clearpage
\begin{deluxetable}{llcccccc}
\tabletypesize{\scriptsize}
\tablewidth{0pt}
\tablecaption{X-ray properties of point sources in young SNRs\label{tab:xps}}
\tablehead{
\colhead{Source} & \colhead{SNR} & \colhead{$d$} & \colhead{Age} &
\colhead{$N_{\rm H}/10^{22}$} & \colhead{$L_{\rm X}\tablenotemark{a}/10^{33}$} & 
\colhead{Adopted Position(s)} &
\colhead{Refs}\\
\colhead{} & \colhead{} & \colhead{(kpc)} & \colhead{(kyr)} & \colhead{(cm$^{-2}$)} & \colhead{(ergs s$^{-1}$)} & &\colhead{} }
\startdata
\pup & Pup A & 2.2 & 3.7 & 0.4 & 3.5 &
08$^{\rm h}$21$^{\rm m}$57\fs42 $-$43\arcdeg00\arcmin16\farcs62
& 1--3  \\
\pks & PKS 1209--52 & 2.1 & 7 & 0.13 & 2.4 & 
12$^{\rm h}$10$^{\rm m}00\fs92$ $-$52\arcdeg26\arcmin28\farcs35\tablenotemark{c} &
2,4,5 \\
 & & & & & & 12$^{\rm h}$10$^{\rm m}00\fs88$
$-$52\arcdeg26\arcmin28\farcs66 & \\ 
\rcw & RCW 103 & 3.3 & 1--3 & 1.7 & 1--60\tablenotemark{b} & 
16$^{\rm h}$17$^{\rm m}36\fs25$ $-$51\arcdeg02\arcmin24\farcs5\phn&6--8 \\
\cas    & Cas A & 3.4 & 0.32 & 1.3 & 2.0 & 
23$^{\rm h}$23$^{\rm m}$27\fs94 $+$58\arcdeg48\arcmin42\farcs51 &
9--12 \\
\enddata
\tablerefs{
(1) \citet{pbw96};
(2) \citet{pst04}
(3) \citet{hb05};
%%%(6) Pavlov et al. 1999.
(4) \citet{hb84};
(5) \citet{dlmc+04};
(6) \citet{tg80};
(7) \citet{psgz02};
(8) \citet{ba02};
(9) \citet{t99};
(10) \citet{pza+00};
(11) \citet{cph+01};
(12) \citet{fps06}.}
\tablenotetext{a}{The X-ray luminosity is in the range 0.5--10 keV for
\pup, 0.4--8 keV for \pks\ and \rcw, and 0.6--6 keV for \cas.}
\tablenotetext{b}{This source shows significant X-ray flux variations
\citep{gpv99}.}
\tablenotetext{c}{The two positions for this object are from
\chandra/ACIS-S and \chandra/HRC-I, respectively; see \S~\ref{subsec:pks}.}
\tablecomments{X-ray positions are all J2000.0, and have typical 90\%
uncertainties of 0\farcs6.}
\end{deluxetable}

%%\clearpage

\begin{deluxetable}{c c c c c c c}
\tablewidth{0pt}
\tabletypesize{\footnotesize}
\tablecaption{Summary of Optical/Infrared Observations of CCOs\label{tab:obs}}
\tablehead{
\colhead{Object} & \colhead{Telescope} & \colhead{Instrument} &
\colhead{Date} & \colhead{Band} & \colhead{Exposure} &
\colhead{Seeing}\\
		 &			&			&
		&		&  \colhead{(min)}	&
\colhead{(arcsec)}\\
}
\startdata
\pup & Magellan Baade 6.5~m & MagIC & 2001-Mar-25 & $r^\prime$ & 10 & 0.7 \\
 & & & & $i^{\prime}$ & 10 & 0.6 \\
 & CTIO Blanco 4~m & OSIRIS & 2002-Feb-25 & $J$ & 9 & 0.8 \\
 & & & & $H$ & 18 & 0.8 \\
 & & & & $K_s$ & 30 & 0.8 \\
 & Magellan Baade 6.5~m & ClassicCam & 2002-Apr-09 & $K_s$ & 30 & 0.6 \\
 & Magellan Clay 6.5~m & MagIC & 2003-Apr-05 & $B$ & 30 & $0.7$ \\
 & & & & $R$ & 20 & 0.7 \\
 & \textit{Spitzer} & IRAC & {2004-Dec-18} & $4.5\ \mu$m & 77 & \nodata \\
 & & & & $8.0\ \mu$m & 75 & \nodata \\
\pks   & Magellan Baade 6.5~m & MagIC & 2001-Mar-24 & $r^\prime$ & 9 & 0.9 \\
 & & & & $i^{\prime}$ & 10 & 1.0 \\
 & & & & $g^{\prime}$ & 10 & 1.0 \\
 & & & 2001-Jun-12 & $u^{\prime}$ & 5 & 1.0 \\
 & CTIO Blanco 4~m & OSIRIS & 2002-Feb-25 & $J$ & 9 & 0.9 \\
 & & & & $H$ & 18 & 0.9 \\
 & & & & $K_s$ & 30 & 0.9 \\
 & Magellan Baade 6.5~m & ClassicCam & 2002-Apr-08 & $J$ & 15 & 0.5 \\
 & Magellan Baade 6.5~m & PANIC & 2004-Feb-12 & $J$ & 30 & 0.5 \\
 &&&& $K_s$ & 40 & 0.5 \\
 & \textit{Spitzer} & IRAC & 2005-Jul-17 & $4.5\ \mu$m & 77 & \nodata \\
 & & & & $8.0\ \mu$m & 75 & \nodata \\
\rcw  & Magellan Baade 6.5~m & ClassicCam & 2002-Apr-09 & $J$ & 10 & 0.6 \\
 &&&& $H$ & 18 & 0.5 \\
 &&&& $K_s$ & 15 & 0.5 \\
& Magellan Clay 6.5~m & MagIC & 2003-Apr-05 & $I$ & 50 & $0.7$ \\
 & \textit{Spitzer} & IRAC & 2005-Mar-29 & $4.5\ \mu$m & 77 & \nodata \\
 & & & & $8.0\ \mu$m & 75 & \nodata \\
\cas &  \textit{Spitzer} & IRAC & 2004-Dec-15 & $4.5\ \mu$m & 77 & \nodata \\
 & & & & $8.0\ \mu$m & 75 & \nodata \\
\enddata
\end{deluxetable}

\begin{deluxetable}{ccccc}
\tabletypesize{\footnotesize}
\tablewidth{0pt}
\tablecaption{Sky brightnesses (in units of MJy/sr) at 
the positions of CCOs \label{tab:sky}}
\tablehead{
\colhead{Filter} & \colhead{\pup} &
\colhead{\pks} & \colhead{\rcw} &
\colhead{\cas} \\}

\startdata
4.5$\mu$ & 0.10  &  0.03     &      3.14  &  0.70 \\
8.0$\mu$ & 4.21  &  1.49     &      11.75 &  81.8 \\
\enddata
\end{deluxetable}

%\clearpage
\begin{deluxetable}{crcccc}
\tabletypesize{\footnotesize}
\tablewidth{0pt}
%\footnotesize
%\centering
%\tablenum{1}
\tablecaption{Magnitudes (detected or 3$\sigma$ limits) of CCOs from
  optical/IR observations \label{tab:lims}}
%\begin{tabular}{cccccc}
%\tableline\tableline
\tablehead{
\colhead{Filter} & \colhead{Wavelength} & \colhead{\pup} &
\colhead{1E~1207.4$-$5209} & \colhead{\rcw} &
\colhead{\cas} \\
 & \colhead{(\AA)} \\
}
 %%%%RX J0852$-$4617\\

\startdata
$u'$ & 3540 & \nodata & $>$22.5 & \nodata      & \nodata \\
$B$  & 4400 & $>$26.5 & \nodata & \nodata      & \nodata \\
$g'$ & 4770 & \nodata & $>$23.9 & \nodata      & \nodata \\
$r'$ & 6230 & $>$25.0 & $>$24.3 & $>$24.8      & \nodata \\
$R$  & 7000 & $>$26.0 & \nodata & \nodata      & \nodata \\
$i'$ & 7620 & $>$24.4 & $>$24.0 & $>$24.4      & \nodata \\
$I$  & 9000 & \nodata & \nodata & $>$24.7      & \nodata \\
%$z'$ & 9130& $>$21.7$& $>$22.5 & $>$22.9      & $>$\\
$J$ & 12500 & $>$21.7 & $>$23.4 & $>$21.9      & \nodata \\   %%$>$21.7 \\
$H$ & 16500 & $>$20.6 & $>$21.3 & 19.4$\pm$0.2 & \nodata \\   %%$>$20.6 \\
$K_s$&21400 & $>$20.1 & $>$22.0 & 18.1$\pm$0.1 & \nodata \\   %%%$>$20.1 \\
$4.5\mu$\tablenotemark{a}&44920&$>$20.7 & $>$20.5 & $>$15.3 & $>$18.7 \\ 
$8.0\mu$\tablenotemark{a}&78700&$>$17.7 & $>$17.6 & $>$13.5 & $>$14.6 \\ 
\enddata
\tablenotetext{a}{The zero magnitude flux are 179.7 Jy and 64.1 Jy for
  the 4.5 and 8.0 $\mu$m bands respectively \citep{rsg+06}.}
%\end{tabular}
\end{deluxetable}

\end{document}